\documentclass[aps,showpacs,preprintnumbers,amsmath,amssymb]{revtex4}
\usepackage{graphicx, color}

\newcommand{\mbb}{\mathbb}

\begin{document}
\title{Coulomb Excitations for a Short Linear Chain of Metallic Shells}
\author {Liubov Zhemchuzhna$^{1,3}$, Godfrey Gumbs$^{1,2}$, Andrii Iurov$^3$, Danhong Huang$^4$ and Bo Gao$^1$}
\affiliation{$^{1}$Department of Physics and Astronomy, Hunter College of the
City University of New York, 695 Park Avenue, New York, NY 10065, USA\\
$^{2}$ Donostia International Physics Center (DIPC),
P de Manuel Lardizabal, 4, 20018 San Sebastian, Basque Country, Spain\\
$^{3}$Center for High Technology Materials, University of New Mexico, NM 87106, USA\\
$^{4}$Air Force Research Laboratory, Space Vehicles Directorate, Kirtland Air Force Base, NM 87117, USA 
}

\begin{abstract}

A self-consistent-field theory is given for the electronic collective modes of a  chain containing
a finite number, $N$, of  Coulomb-coupled spherical two-dimensional electron gases (S2DEs)
arranged with their centers along a straight line, simulating a narrow micro-ribbon 
of metallic shells.     The separation 
between nearest-neighbor shells is arbitrary and because of the quantization of the electron 
energy levels due to their confinement to the spherical surface, all angular momenta 
$L$ of the Coulomb excitations and their projections $M$ on the quantization axis are coupled.
However,  for incoming light with a specific polarization, only one angular momentum quantum number 
is chosen. We  show that when $N=3$  the next-nearest-neighbor Coulomb coupling is larger than 
its value if they are located at opposite ends of a right-angle triangle forming the 
triad. Additionally,   the frequencies of the plasma excitations depend on the 
orientation of the line joining them with respect to the axis of quantization 
since the magnetic field generated
from the induced oscillating electric dipole moment on one
sphere can couple to the induced magnetic dipole moment on another.

\end{abstract}


\maketitle

\section{  Introduction}
\label{sec1}

Over the years, plasmon excitations have been investigated both experimentally
and theoretically for topologically different nanostructures such as the
two-dimensional electron gas (2DEG) and for systems with layers of 2DEG 
forming a layered electron gas (LEG) systems \cite{AFS,Jain-Allen}.  The LEG may be obtained  in a
multilayer semiconductor system by epitaxial growth such as GaAs/AlGaAs or in a type-II system
such as GaSb AlSb InAs quantum-well structures
 containing an electron and a hole gas layer in each unit cell \cite{Xiaodong}.
Another class of nanostructures for which the Coulomb excitations have been studied 
is the spherical 2DEG (S2DEG) \cite{Inaoka,Devreese,Yannouleas}  which  has been employed in 
modeling  metallic 
dimers \cite{2spheres,Nordlander},   clusters of carbon nanoparticles \cite{3spheres}  and 
metallic chains  of gold nanoparticles  \cite{infinite,Reviewer_1-1}. 
However, since the spherical geometry allows for possible anisotropic coupling  
(unlike the cylindrical geometry \cite{aizin,mcneish})
with an external light source by employing  far-field polarization
spectroscopy \cite{Reviewer_1-1}, the Coulomb excitations for the S2DEG  allow
for more variation in frequency than the 2DEG by employing shells of different radii, 
forming finite length chains with adjustable nearest-neighbor separations, or 
using bundles arranged in assorted configurations.   Furthermore, we may model the S2DEG 
as an incompressible fluid  \cite{Nordlander} or as a point dipole  \cite{Reviewer_1-2} 
 in which only nearest-neighbor  dipole  interactions through an EM field 
 are assumed dominant  \cite{Reviewer_1-2}.  But, in such a model,
the dipoles have been assumed  to point in a chosen direction (like in an antenna array)
  so that only one component of  the angular momentum  is employed, e.g., if $L=1$ then 
	only one of $M=0,\pm 1$ plays a role. In this paper, we overcome this restriction 
in investigating  longitudinal and transverse plasmon-polariton modes in a chain with a finite number $N$
of metallic shells. Although our formalism may be applied to arbitrary $N$, we present results when $N=3$ 
since it affords us with the opportunity of comparing with the configuration when the 
shells are positioned with their centers at the vertices of a right-angle triangle. This example
demonstrates the  orientational dependence of the Coulomb matrix elements as well as the effects
arising from the finite  length of a linear chain.

\par
\medskip

Our model  described below is drastically different from that in  Ref. \cite{Reviewer_1-2}
 because we include dynamically  screened  long-range Coulomb interactions and the  
orbital angular momenta and their projections onto the quantization
axis   are coupled. This means that the  localized plasmons on each shell are 
coupled by the inter-sphere electron-electron Coulomb interaction.
Also, an  important difference between the present model and the point-dipole model 
is that the electromagnetic field generated by time-varying charge
distributions or current  is retarded, whereas in our model
the inter-shell coupling is electrostatic.  Since the polarization
function for the S2DEG vanishes for $L=0$, there are no plasmon excitations
for $L=0$ \cite{Inaoka}. This is in contrast with the model in Ref. [\onlinecite{Reviewer_1-1}]
which allows plasmon excitations when the dipoles have a fixed direction,
 i.e.,   which corresponds to a chosen projection for  an unspecified  angular momentum.

\begin{figure}[!ht]
\centering
\includegraphics[width=0.48\textwidth]{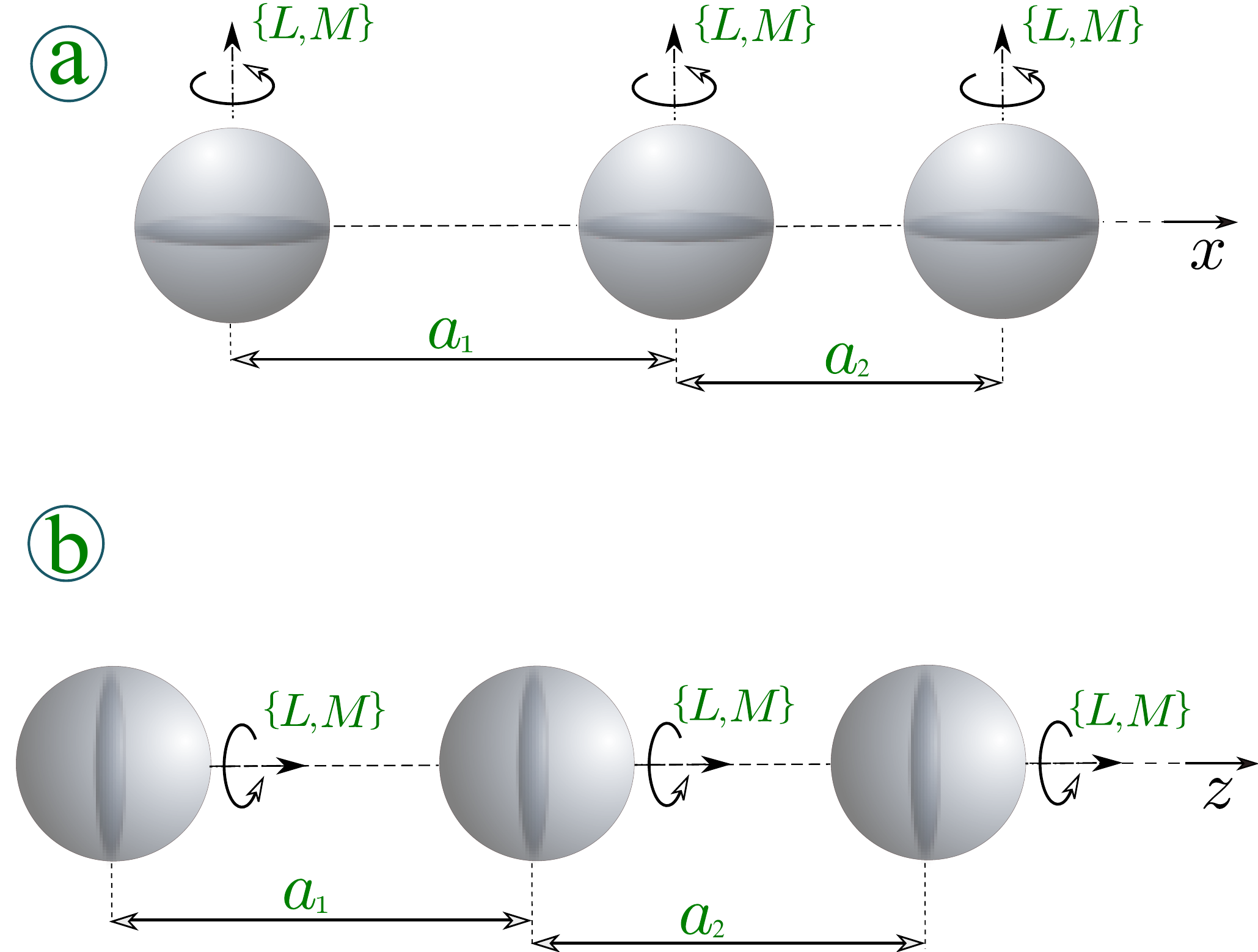}
\caption{(Color online)  Schematic illustration of a linear triad of S2DEGs.
 The angular momentum is $L$ and its component is $M$. The shells are aligned along the 
x-axis. We consider two different cases of the quantization axis direction - along 
the x-axis (plot $(a)$) and along z-axis (plot $(b)$). The middle shell is centered 
in the origin,  the distances between the centers are $a_1$ and $a_2$, as indicated.    }
\label{FIG:1}
\end{figure}
%
%-----
 
\medskip
\par

With the use of our formalism, we may directly simulate the  modes
of oscillation   observable by employing circularly-polarized
light or a helical light beam. That is,
the plasmons are detected experimentally using an external polarizing field, which specifies a preferred
direction, which is the $z-$axis in our case. Our primary goal is to consider a closely
packed finite-length assembly of S2DEGs with all their centers  on a straight line relative to the 
quantization axis. In such a system, the strength of the  interaction between different nano shells varies
with  their separation as well as the  relative orientation of the line joining their centers. 
With this in mind, it makes sense to distinguish the present  investigation     
from both cases of a   triad arranged at the vertices of a right-angle triangle \cite{3spheres} 
or an infinite linear array of periodically placed S2DEGs \cite{infinite}. The assembly under 
consideration is shown schematically  in Fig. \ \ref{FIG:1}.
We refer to each shell as $j=\,1,\,2,\,3, \cdots,\  N$, starting  with the one on the far left. 
For $N=3$, the  middle shell is  centered at the origin. We specify the distances between the 
S2DEGs as $a_{12} = a_1$ and $a_{23}= a_2$, so that $a_{13}=a_1+a_2$. In all our  numerical
considerations, we investigate   separately the cases corresponding to 
equal and unequal distances $a_1$ and $a_2$ in conjunction with equal and unequal 
radii of the shells.  We consider 
either the case of three identical shells with $R_1=R_2=R_3$ or the situation when the middle 
shell is larger than the other two: $R_2 > R_1 = R_3$.      
  
\medskip
\par

The spatial anisotropy of the plasmon excitations is just another example of 
the orientational dependence of the optical and transport properties of condensed matter systems. 
We now cite a few examples when such an anisotropy has  been demonstrated. These include
the  electrical, thermal, mechanical and chemical properties of graphite along the $a,\ b$ and $c$ directions
\cite{n21} and the  elastic properties of carbon nanotube bundles \cite{n22}. Also, the dispersion
relation of the high-frequency optical $\pi$-plasmons for graphite was calculated by Chiu, et al., \cite{n23} 
and it was shown that the plasmon excitations depend on whether the momentum transfer is parallel or perpendicular
to the hexagonal plane lying in the Brillouin zone. The anisotropic conductivity of epitaxial 
graphene on SiC was reported in Ref. \cite{n24}.
We note that there have been reports in which some of these anisotropic properties  have
been used in device applications. For example, in Ref. \cite{n42},  
the tuning of surface plasmon frequencies to more efficient optical sensors was investigated.

\medskip

The rest of this paper is presented  in the following way. In Sec.\ \ref{sec2} 
 the theoretical formulation for deriving the plasmon equation for three spherical shells
whose centers are on a straight line. The Coulomb matrix elements are shown to depend on the relative 
orientation of this line-of-centers with respect to the axis of quantization for the 
angular momentum associated with the spherical geometry. The plasmon equation in 
general involves  the coupling of all angular momenta and their projections on the quantization axis. 
But, in Sec.\ \ref{sec3}, we obtain the dielectric matrix assuming the impinging
light sourse probing the Coulomb excitations has a specified polarization. Section\ \ref{sec4}
is devoted to a presentation and discussion of our numerical results. We conclude with a summary of
our results in Sec.\ \ref{sec5}.

\section{Theoretical Formulation of the Problem}
\label{sec2}

We turn our attention to a system of $N$ spherical shells with their
centers on the $x$ axis.  The center of one of the spheres is located
at $x=0$ with radius $R_0$ whereas the other spheres are centered
at $x=--a_1$ and its radius is   $R_1$,  $x=a_2$ and radius $R_2$,
and so on. We assume no overlap of the shells so that we demand, for example, that the inequalities
 $a_1>R_1+R_0$ and $a_2>a_0+a_2$ are satisfied. In the absence of tunneling
between the shells, the wave function for an electron on the
$j$-th shell ($j=1,2,3, \ \cdots \ , \ N$) is given by

\begin{equation}
<{\bf r}\mid j\nu>=\Psi_{jlm}\left(\vec{r}-(j-1)a\hat{e}_x\right),
\hspace{.3cm}
\Psi_{jlm}(\vec{r})=f_j(r)\frac{1}{\sqrt{ R_j^2}}
 Y_{lm}(\Omega)    \ ,
\label{eq1.37}
\end{equation}
with $\nu=\{l,m\}$ and $f_j^2(r)=\delta(r-R_j)$. The
energy spectrum has the form

\begin{equation}
\epsilon_{j,\nu}=\frac{\hbar^2l(l+1)}{2m^{\ast}R_j^2}\;.
\label{eq:sp}
\end{equation}
The equation of motion of the density matrix operator is

\begin{equation}\label{eq1.39}
\imath\hbar\frac{\partial\hat{\varrho}}{\partial
t}=[\hat{H},\hat{\varrho}] \ ,
\end{equation}
where $\hat{H}=\hat{H}_0-e\Phi$ is the Hamiltonian of the electron
on the surface of the sphere, $\hat{H}_0$ is the free electron
Hamiltonian and $\Phi$ is the induced potential. The potential
$\Phi$ satisfies Poisson's equation

\begin{equation}
\nabla^2\Phi({\bf r},\omega)=\frac{4\pi e}{\varepsilon_s}
\delta n({\bf r}, \omega) \ ,
\label{eq1.40}
\end{equation}
where $\varepsilon_s\equiv 4\pi\epsilon_0\varepsilon_b$
 and $\varepsilon_b$ is the uniform background dielectric
constant. Additionally,  $\delta n{(\bf r),\omega}$ is the induced electron
density. We use linear response theory to calculate the
induced particle density as

\begin{equation}
\delta n({\bf r},\omega)=\sum\limits_{j,j\prime}
\sum\limits_{\nu,\nu^\prime} <{\bf r}\mid j\nu>
<j\nu\mid\hat{\varrho}_{1}({\bf r},\omega)\mid j^{\prime}
\nu^\prime><j^{\prime}\nu^{\prime}\mid{\bf r}> \ ,
\label{eq1.41}
\end{equation}
where

\begin{equation}
<j\nu\mid\hat{\varrho}_1({\bf r},\omega)\mid
j^{\prime}\nu^\prime>=2e\,\frac{f_0(\epsilon_{j\nu})-
f_0(\epsilon_{j^{\prime}\nu^\prime})}{\hbar\omega
+\epsilon_{j^{\prime}\nu^\prime}-\epsilon_{j\nu}}
<j\nu\mid\Phi({\bf r},\omega)\mid
j^{\prime}\nu^\prime> \ ,\label{eq1.42}
\end{equation}
in terms of the Fermi-Dirac distribution function $f_0(\epsilon)$ and we express  
the induced potential as $\Phi({\bf
r})=\frac{1}{V_0}\sum\limits_{\bf q^\prime} \Phi({\bf q^\prime})
e^{i{\bf q^\prime}\cdot{\bf r}}$, where $V_0$ is a normalization volume. 
Then Eq.~(\ref{eq1.41}) becomes

\begin{eqnarray}
\delta n({\bf r},\omega)&=&\frac{2e}{V_0}\sum\limits_{j,j^\prime}
\sum\limits_{\nu,\nu^\prime} \frac{f_0(\epsilon_{j\nu})-
f_0(\epsilon_{j^{\prime}\nu^\prime})}
{\hbar\omega+\epsilon_{j^{\prime}\nu^\prime}-\epsilon_{j\nu}}
<{\bf r}\mid j\nu><j^\prime\nu^\prime\mid{\bf r}>\nonumber\\
&\times&\sum\limits_{{\bf q^\prime}}\Phi({\bf q^\prime})<j\nu\mid
e^{i{\bf q^\prime\cdot r}}\mid j^\prime\nu^\prime> \ ,
\label{eq1.43}
\end{eqnarray}
or by taking the Fourier transform with respect to ${\bf r}$

\begin{eqnarray}
\delta n({\bf q},\omega)&=&\frac{2e}{V_0}\sum\limits_{j,j^\prime}
\sum\limits_{\nu,\nu^\prime} \frac{f_0(\epsilon_{j\nu})-
f_0(\epsilon_{j^{\prime}\nu^\prime})}
{\hbar\omega+\epsilon_{j^{\prime}\nu^\prime}-\epsilon_{j\nu}}
<j^\prime\nu^\prime\mid e^{-i{\bf q}\cdot{\bf r}}\mid j\nu>
\nonumber\\
&\times&\sum\limits_{\bf q^\prime}\Phi({\bf q^\prime})
<j\nu\mid e^{i\bf q^\prime\cdot r}\mid j^\prime\nu^\prime>\;.
\label{eq1.44}
\end{eqnarray}
The matrix elements $<j\nu\mid e^{i\bf q\cdot r}\mid
j^\prime\nu^\prime>$ with wave functions $<{\bf r}\mid j\nu>$ given
by Eq.~(\ref{eq1.37}) may be calculated using the
expansion of a plane wave in spherical waves

\begin{equation}
e^{i\textbf{q}\cdot \textbf{r}\ }=4\pi\sum_{l,m} \ \ i^l
j_l(qr)  Y_{lm}^\ast(\hat{\textbf{q}})
Y_{lm}(\hat{\textbf{r}})
\label{eq1.45}
\end{equation}
where $j_l(x)$ is a spherical Bessel function.
The result is

\begin{eqnarray}
&& <j\nu\mid e^{i\bf q\cdot r}\mid j^\prime\nu^\prime>
=4\pi \delta_{jj^\prime}
e^{i(j-1)q_xa}
\nonumber\\
&\times&\sum_{L,M} \ \ i^L j_L(qR_j)   Y_{LM}^\ast(\hat{\textbf{q}})
\int d\Omega\ \  Y^{\ast}_{lm}(\Omega)
Y_{LM}(\Omega)\
Y_{l^{\prime}m^{\prime}}(\Omega)
\;,
\label{eq1.46}
\end{eqnarray}

 Substituting Eq.~(\ref{eq1.46}) into Eq.~(\ref{eq1.44}), we obtain after some
algebra

\begin{eqnarray}
&& \delta n({\bf q},\omega)=(4\pi)^2\frac{2e}{V_0}\
\sum_{l,m}
\sum_{ l^\prime, l^\prime}
\sum_{j=1,2}
\frac{f_0(\epsilon_{j,l})-f_0(\epsilon_{j,l^\prime})}
{\hbar\omega+\epsilon_{j,l^\prime}-\epsilon_{j,l}}
\  e^{-i(j-1)q_xa}
\nonumber\\
&\times&\sum_{L,M}\  \ (-i)^L j_L(qR_j)
Y_{LM}(\hat{\textbf{q}})\
\int d\Omega\ \  Y^{\ast}_{l^\prime m^\prime}(\Omega)
Y_{LM}^\ast(\Omega)\
Y_{lm}(\Omega)
\nonumber\\
&\times& \sum_{q_x^\prime, q_y^\prime , q_z^\prime} \
e^{i(j-1)q_x^\prime a}
\Phi\left(q_x^\prime,q_y^\prime,q_z\right)
\nonumber\\
&\times&\sum_{L^\prime,M^\prime} \    i^{L^\prime}
j_{L^\prime}(q^\prime R_j)
Y_{L^\prime M^\prime}^\ast(\hat{\textbf{q}}^\prime)
\int d\Omega\ \  Y^{\ast}_{lm}(\Omega)
Y_{L^{\prime}M^{\prime}}(\Omega)\
Y_{l^{\prime}m^{\prime}}(\Omega)
 \ .
\label{eq1.47}
\end{eqnarray}
or

\begin{eqnarray}
&& \delta n({\bf q},\omega)=(4\pi)^2\frac{2e}{V_0}\sum_{j=1,2}\sum_{L}
\sum_{l,l^\prime}
\frac{f_0(\epsilon_{j,l})-f_0(\epsilon_{j,l^\prime})}
{\hbar\omega+\epsilon_{j,l^\prime}-\epsilon_{j,l}}
\ (2l+1)(2l^\prime+1)
\left( \begin{matrix}
l&l^\prime& L\cr
 0 & 0 & 0\cr
\end{matrix}\right)^2
\nonumber\\
&\times&   e^{-i(j-1)q_xa} \sum_{M}\  \   j_L(qR_j)
Y_{LM}(\hat{\textbf{q}})\
\nonumber\\
&\times& \sum_{q_x^\prime, q_y^\prime , q_z^\prime} \
e^{i(j-1)q_x^\prime a}
\Phi\left(q_x^\prime,q_y^\prime,q_z\right)
j_{L}(q^\prime R_j)
Y_{L M}^\ast(\hat{\textbf{q}}^\prime)
 \ .
\label{eq1.47+}
\end{eqnarray}
in terms of Clebsch-Gordon coefficients.
Taking the Fourier transform of Eq.~(\ref{eq1.40}) we have
$\Phi({\bf q})=-4\pi e\delta n({\bf q})/ \varepsilon_s q^2$. Using
this relation in Eq.~(\ref{eq1.47}), we obtain

\begin{equation}
\delta n({\bf q},\omega)=-\frac{8\pi e^2}{\varepsilon_s} (4\pi)^2 \sum_{j,L,M}\Pi_{j,L}(\omega)
e^{-i(j-1)q_xa}  \   j_L\left(q R_j \right)U_{j,L,M}
Y_{LM}(\hat{\textbf{q}})\ ,
\label{eq1.48}
\end{equation}
where $\Pi_{j,L}(\omega)$ is the density response function of
the $j$-th nano shell   with

where

\begin{equation}
\Pi_{L}(\omega)=  \sum_{l,l^\prime}
\frac{f_0(\epsilon_{l})-f_0(\epsilon_{l^\prime})}
{\hbar\omega+\epsilon_{l^\prime}-\epsilon_{l}}
(2l+1)(2l^\prime+1)
\left( \begin{matrix}
l&l^\prime& L\cr
 0 & 0 & 0\cr
\end{matrix}\right)
^2
\label{gae13}
\end{equation}
and

\begin{equation}
U_{j,LM} = \frac{1}{L_xL_yL_z}\sum_{q_x,q_y,q_z}e^{i(j-1)q_xa}
\frac{\delta n(q_x,q_y,q_z,\omega)}{q_x^2+q_y^2+q_z^2}
\ j_L\left(q R_j\right)
 Y_{LM}^\ast(\hat{\textbf{q}}) \ .
\label{eq1.49}
\end{equation}
Here, $L_x,L_y$ and $L_z$ are normalization lengths, with $V_0=L_xL_yL_z$. Substituting the
expression for $\delta n({\bf q})$ given in Eq.~(\ref{eq1.48}) into
Eq.~(\ref{eq1.49}), we obtain

\begin{equation}
\sum_{j^\prime=1}^N\sum_{L^\prime=0}^{\infty}
\sum_{M^\prime=-L^\prime}^{L^\prime}
\left[\delta_{jj^\prime}\delta_{LL^\prime}\delta_{MM^\prime}+
\frac{2e^2}{\varepsilon_s}\Pi_{j^\prime,L^\prime }(\omega)
V_{j^\prime L^\prime M^\prime,jLM}(R_j,R_{j^\prime},a)\right]
U_{j^\prime,L^\prime M^\prime}\  =0 \ ,
\label{eq1.50}
\end{equation}
where

\begin{eqnarray}
V_{j^\prime L^\prime M^\prime,jLM}(R_j,R_{j^\prime};a)
&=&8\int
\frac{d^3\textbf{q}}{q^2}\
  j_L(qR_j)j_{L^\prime}(q R_{j^\prime})
  Y_{LM}^\ast(\hat{\textbf{q}}) Y_{L^\prime M^\prime}(\hat{\textbf{q}})
e^{i(j-j^\prime)q_xa}
\nonumber\\
&=& \frac{\pi}{2R(2L+1)}\delta_{L,L^\prime}\delta_{M,M^\prime}
\  \  \  \mbox{when}\ \ j=j^\prime
\ .
\label{eq1.51}
\end{eqnarray}
We have explicitly by setting  $j=1,2$ or $j=3$ in turn for each of
the three spheres

\begin{eqnarray}
 & & \left[1+\frac{2e^2}{\varepsilon_s (2L+1)R_j}
 \Pi_{j,LM}(\omega) \right] U_{j,LM}
 \nonumber\\
&+& \frac{2e^2}{\epsilon_s} \sum_{j^\prime\neq j}\sum_{L^\prime, M^\prime}
\Pi_{j^\prime,L^\prime }(\omega)
V_{j^\prime L^\prime M^\prime,jLM}(R_j,R_{j^\prime},a)
U_{j^\prime,L^\prime M^\prime}\  =0 \ ,
\label{eq1.51b}
\end{eqnarray}
If the spheres are identical, then we
need only set $j=1$, but still have to do the sum over $j^\prime=1,2,3$.
The set of linear equations (\ref{eq1.51b})  have
nontrivial solutions provided the determinant of the coefficient
matrix of $\{ U_{j,LM}   \} $ is zero.  Consequently, plasmon
 modes with different values of $L $ on the two shells may now be
coupled via the Coulomb interaction. Since
$V_{1LM, L^\prime M^\prime}(R_1,R_2;a)\to 0$ in the limit
$a\to\infty$, this matrix is diagonal when $a\gg R_1,R_2$ and the
plasmon mode  equation reduces to the result for isolated shells

\begin{figure}
\centering
\includegraphics[width=0.35\textwidth]{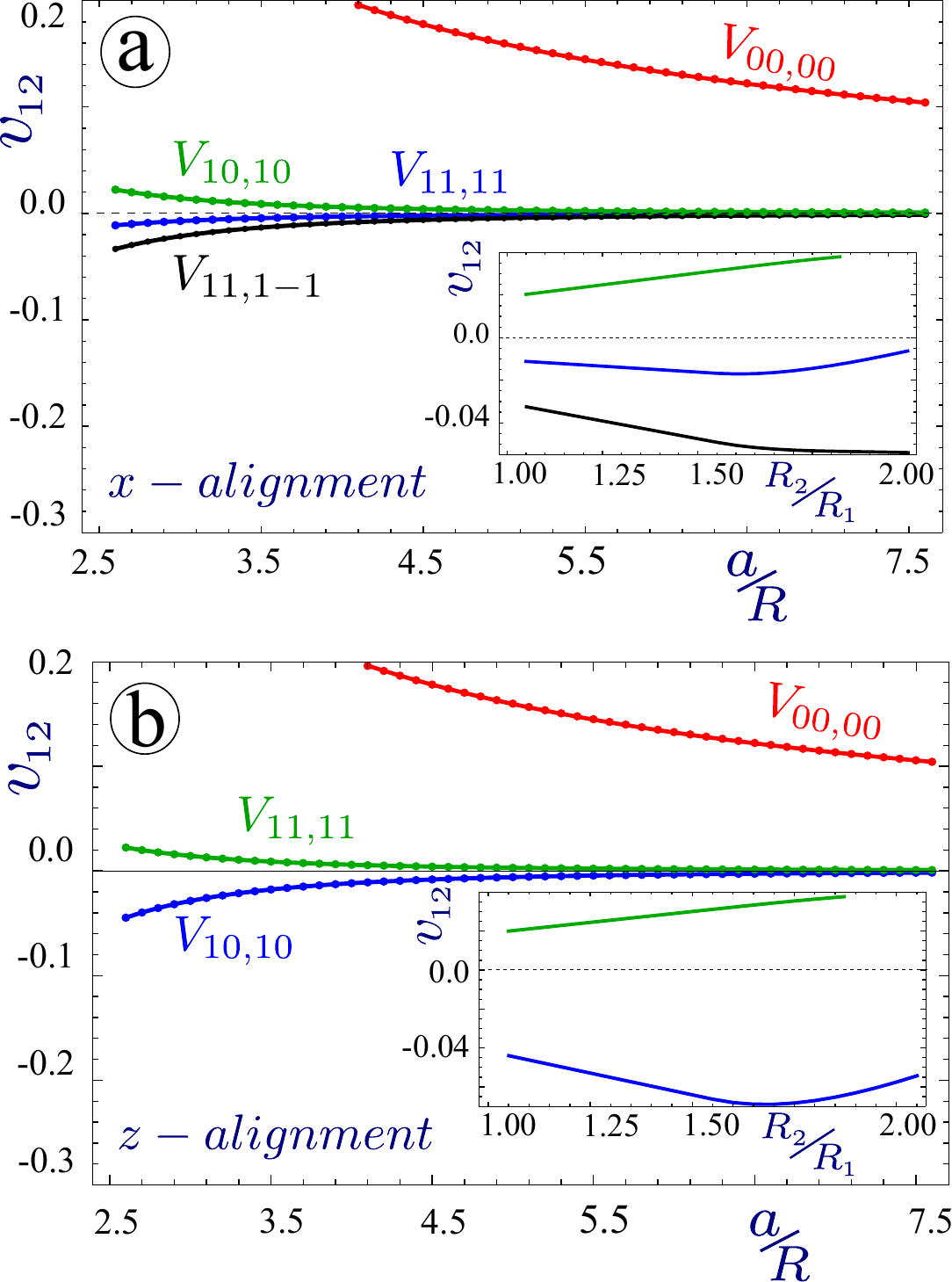}
\caption{(Color online)  Coulomb matrix elements $v_{12}$ (in units of $2 e^2/(\pi \epsilon_s R) = 5.7\,eV$ ) 
for a pair of  interacting  S2DEGs  as a function of the separation 
$a$ (in units of their radius  $R=1\,nm$) between their centers.  The displayed matrix elements are 
$V_{00,00}$ (red curves),   $V_{10,10}$ (green curves),  $V_{11,11}$ (blue curves),
and $V_{11,1-1}$ (black curves).  Plot $(a)$ shows the case of $x-$alignment, and plot $(b)$ 
shows results for $z-$alignment. The largest matrix element, $V_{00,00}$, which corresponds 
to $L=L^\prime=0$ and $M=M^\prime=0$ does not contribute to the plasmon excitation spectrum. However,
it  is given for comparison. The insets demonstrate the corresponding interaction potential matrix
elements  (same color)  as a function of the ratio  $R_2/R_1$ of the radii of the two shells for fixed separation
$a$  between their centers.    
}
\label{FIG:2}
\end{figure}
%
%-----

\begin{figure}
\centering
\includegraphics[width=0.52\textwidth]{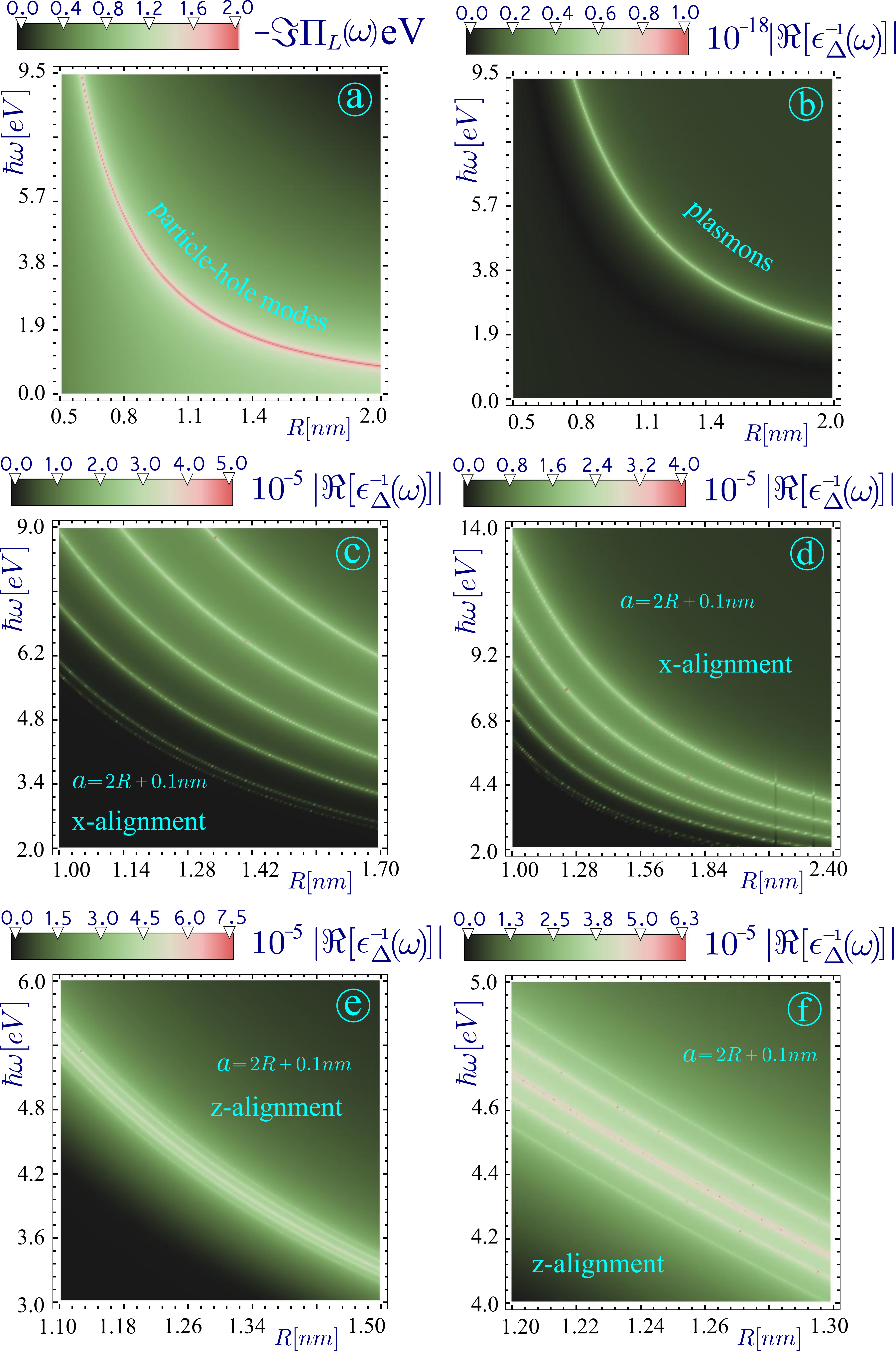}
\caption{(Color online)  Particle-hole modes and plasmon excitations for a
single S2DEG and for a linear assembly of three shells as a function
of their radii. Panel $(a)$ shows the particle-hole modes for a single
S2DEG. Panel $(b)$ presents the plasmon excitation frequency of  a single S2DEG
as a function of  its radius. All the other plots in panels $(c)$ through $(f)$  
demonstrate the  dependence of the plasmon frequency as a function of radius $R$ 
for three equivalent S2DEGs with $R_1=R_2=R$ and with equal separations $a_1=a_2=a=2R+0.1\,nm$. Panels
$(c)$ and $(d)$ represent the plasmon branches of an $x-$aligned linear 
triad with various ranges of their radii, i.e., from $1.0\, \mbox{to} \, 1.7\, nm$ for 
$(c)$ and $1.0\, \mbox{to} \, 2.4\, nm$ for $(d)$. Plots in $(e)$ and $(f)$ show the
corresponding plasmon dependence for the case of $z-$alignment. While plot
$(e)$ shows results for radii in the range  from $1.1\, \mbox{to} \, 1.5\, nm$, 
and plot $(f)$ gives 
the dependence over a narrower range $1.2\, \mbox{to} \, 1.3\, nm$.         } 
\label{FIG:3}
\end{figure}
%
%-----

\begin{equation}
\prod_{L}\varepsilon_{1,L}^{2L+1}(\omega)
\varepsilon_{2,L}^{2L+1}(\omega)  \varepsilon_{3,L}^{2L+1}(\omega)=0 \ ,
\label{eq1.54}
\end{equation}
where $\varepsilon_{j,L}(\omega)$ is the dielectric function for
the $j$-th shell. The significance of equations  (\ref{eq1.51b})
for chosen $L,M$  is that they give explicitly the effect of
the Coulomb interaction on  each  shell through $\varepsilon_{j,L}$  as
well as the coupling  between the pair of shells through the
Coulomb matrix elements $V_{j^\prime L^\prime M^\prime,jLM}(R_j,R_{j^\prime},a)$.
Additionally, the nature of this coupling may be characterized in the following
way when carrying out numerical calculations. For chosen $L$     and
$M$ satisfying $-L\leq M\leq L$, there are $2(2L+1)\times 2(2L+1)\times 2(2L+1)$
elements in a block
sub-matrix which includes $(2L+1)$ elements along the diagonal,   equal to
$\varepsilon_{1,L} (\omega)$,  $(2L+1)$ diagonal elements
   equal to $\varepsilon_{2,L} (\omega)$  and $(2L+1)$ diagonal elements
   equal to $\varepsilon_{3,L} (\omega)$. For example, if we
   consider the coupling between
   sub-matrices with angular momentum $L=1,2,3,\cdots,N$, then
   the dimension of the matrix is $3\sum_{L=1}^N(2L+1)=3N^2+6N$.
   Specifically, if we use just the $L=1$ sub-matrix, we have
   a $9\times 9$ matrix which we discuss below.

%%%%%%%%%%%%%%%%%%%%%%%%%%%%%%%%%%%%%%%     XXXXXXXXXXX

\section{Dielectric function  matrix    calculation}
\label{sec3}
%

%%%% \textcolor{red}{We start modifying the text from here:} 

\medskip
\par

We now restrict our attention to the case when  $N=3$  and the angular momentum quantum
number  is chosen as  $L=1$. The zeros of the dielectric function for the three aligned S2DEGs, which determine 
 the plasmon excitation energies of the system, may be obtained as the solution of a 
determinantal equation Det $\tensor{ {\cal M}}_{\epsilon} = 0$, where $\tensor{ {\cal M}}_{\epsilon}$
is a complete $9 \times 9$ matrix, including the dielectric function of each nanosphere, 
as well as the inter-sphere interaction potential elements. In our formalism above, we 
took the spherical shells as being centered on the $x$-axis. However,  without loss of
generality,  the spheres may be placed along the axis of quantization, the $z$-axis.
In either case, the dielectric matrix may be presented as a block matrix as follows:

\begin{equation}
\tensor{{\cal M}}_{\epsilon} = \left({ \begin{array}{ccc}
{\cal \underline{D}}_1 & \mbb{V}^{A}_{12} & \mbb{V}^{S}_{13} \\
\mbb{V}^{A}_{21} & {\cal \underline{D}}_2 & \mbb{V}^{A}_{23} \\
\mbb{V}^{S}_{31} & \mbb{V}^{A}_{32} & {\cal \underline{D}}_3 															
\end{array}  }\right)\ ,
\label{ma}
\end{equation}
where, in our notation, the submatrix ${\cal \underline{D}}_i$ is diagonal.
   Each diagonal block consists of three identical elements, representing the
dielectric function of an isolated S2DEG: We have

\begin{equation}
{\cal \underline{D}}_i=\left({  \begin{array}{ccc}
\epsilon_{L=1}(R_i,\,\omega) & 0 & 0 \\
0 & \epsilon_{L=1}(R_i,\,\omega) & 0\\
0 & 0 & \epsilon_{L=1}(R_i,\,\omega)\end{array} }\right)\ ,
\label{eqa2}
\end{equation}
for $i=1,2$ or $3$. The coupling between  different shells is given by the off-diagonal 
blocks $\mbb{V}^{A}_{jj'}$ and $\mbb{V}^{S}_{jj'}$, representing the interaction
potentials between the adjacent $(1-2)$ and $(2-3)$ shells, and  next-nearest-neighbors 
$(1-3)$. This geometrical arrangement leads to the main specific feature of our 
case of the aligned shells.  In this regard, we  
  consider an  off-diagonal $3\times3$ interaction matrix $\mbb{V}_{jj'}$, which 
may denote either $\mbb{V}^{A}_{jj}$ or $\mbb{V}^{S}_{jj}$ and include their common features
as demonstrated in Fig.\ \ref{FIG:1}.
We emphasize that the analytical formula of the matrix and the expression for each elements 
are identical for a next-nearest-neighbor  or an adjacent pair of S2DEG's, while the only difference comes from
the distance between the shells, resulting in the value $a_{12}+a_{23}$ for the next-nearest-neighbor
 pair. Specifically,  for $L=L'=1$, the matrix consists of \textit{nine} elements determined by the $M$ and $M'$ as follows: 

\begin{equation}
\label{elz2}
 \mbb{V}_{jj'}^{M\,M'} ( R; a_{jj'})=\Pi_{L=1}(R,\,\omega)
\left({  \begin{array}{ccc}
V_{M=-1,M'= -1}  & V_{M=-1,M'=0}   & V_{M=-1,M'=1}\\
V_{M=0,M'= -1}  &  V_{M=0,M'=0}  &  V_{M=0,M'=1}\\
V_{M=1,M'= -1}  &  V_{M=1, M'=0}  & V_{M=1,M'=1}
\end{array}  }\right)\ .
\end{equation}
Each matrix elements depends on the orientation of the nanoshells and will be calculated below
according to Eq.(\ref{eq1.51}). 
\par

\begin{figure}
\centering
\includegraphics[width=0.48\textwidth]{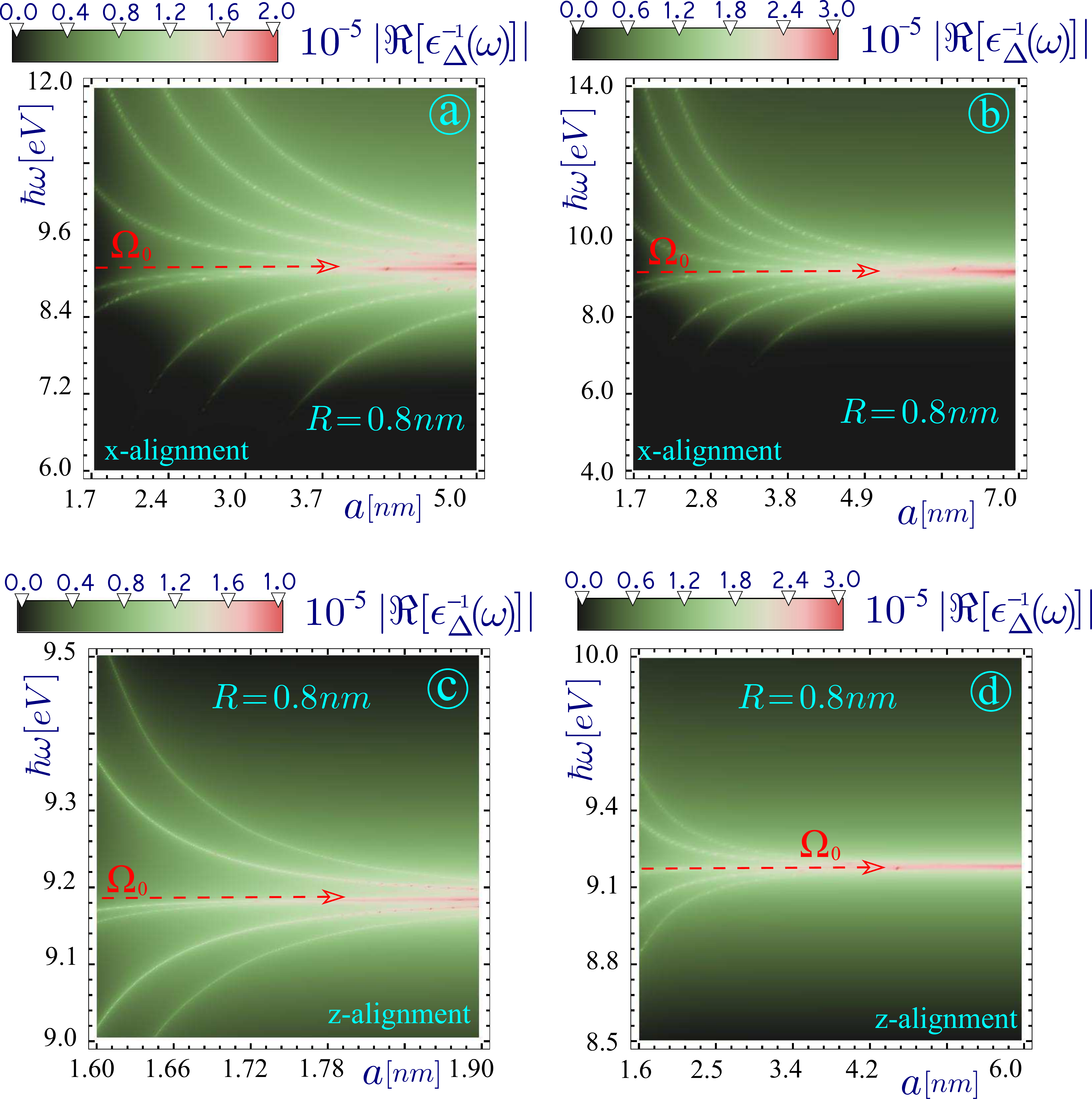}
\caption{(Color online) Plasmon excitations for a linear assembly of three
identical S2DEGs as a function of the separation $a$ ($a_1=a_2=a$). The radius
of the spheres was chosen to be $R=R_1=R_2=0.8\,nm$. The upper panels $(a)$ and $(b)$
correspond to the case of  $x-$alignment, and the lower ones $(c)$ and $(d)$   to
$z-$alignment. Each panel presents results for a  specific range of the separation $a$:  
$1.7\, \mbox{to} \, 5.0\, nm$ for $(a)$ and $(b)$, respectively.   In panel $(c)$,
the range is $1.6\, \mbox{to} \, 1.9\, nm$ and       $1.6\, \mbox{to} \, 6.0\, nm$   in panel $(d)$. 
The solution for a system of non-interacting shells $\Omega_0=9.17\,eV/\hbar$ is  
 indicated for comparison. }
\label{FIG:4}
\end{figure}

\medskip
\par
Each matrix element in Eq.\ (\ref{eqa2}) may be expressed as  
$\epsilon_{L=1}(R_i,\,\omega) = 1+2e^2/(3\epsilon_s R)\,\Pi_{L=1}(\omega)$. 
Consequently, for the simplest case of three  non-interacting shells of equal 
radius $R$, the plasmon  branches are obtained by solving for the zeros of the function

\begin{equation}
{\mbox Det}\  \tensor{{\cal M}}_{\epsilon} = 
({\mbox Det}\  \underline{{\cal D}})^3 = (1+2e^2/(3\epsilon_s R)\,\Pi_{L=1}(\omega))^9  
\end{equation}
i.e. we obtain a  ninefold degenerate  plasmon solution for a single isolated S2DEG. 
This solution corresponds to the result for  interacting spheres as a function of their 
separation when both  $a_{12}/R_i \to \infty$ and $a_{23}/R_i \to \infty$.
In proceeding to take the inter-shell Coulomb coupling into account, we must distinguish between
 the  $z-$ and $x-$alignments, leading to  different analytical forms for the interaction submatrix. 
Let us now consider the interaction matrix elements for both cases of alignment.
We conclude that the result of the integration is non-zero only for certain    
elements, representing an interesting set of \textit{selection rules}. The situations varies 
for each   alignment. For   $x-$alignment, the order of the remaining 
Bessel function is determined by the difference $\vert M - M'\vert$ and the elements are
classified correspondingly (see Eq. \ (\ref{elx})). For example, all the elements with 
$\vert M - M'\vert$ = 1 are equal to zero due to the inherent symmetry of the polar integral.
For  $z-$alignment, the azimuthal angle $\phi-$ is determined   by the phases of the 
spherical harmonics in Eq.\ (\ref{eq1.51}), so that only elements with $M = M'$ are non-zero, i.e.
we obtain only two such elements. 

 \medskip
\par

We first consider  the simplest case of three identical shells each of  radius $R$ and 
with the same number of filled energy levels or Fermi level $L_F$. We set
the separation $a_{12} = a_{23} = a$, so that $a_{13} = 2a$. In previous work, we obtained 
\cite{2spheres} the pair-wise interaction potential matrix for two
S2GEGS with radii $R_1$ and $R_2$,  aligned along the z-axis with separation $\alpha$ 

\begin{equation}
\label{elz}
 \mbb{V}_{jj'}=\Pi_{L=1}(R,\,\omega)
\left({  \begin{array}{ccc}
v_1  &  0  & 0\\
0  &  v_3  &  0\\
0  &  0  & v_1
\end{array}  }\right)\ ,
\end{equation}
where

\begin{eqnarray}
 && v_1 = V_{11,11} =\frac{6 e^2}{\pi \epsilon_s} \int\limits_{0}^{\infty} 
 \frac{dq}{(\alpha q)^3}\left\{\sin(\alpha q) - \alpha q \cos(\alpha q) \right\} j_1(R_1 q)j_1(R_2 q)  
 \nonumber\\
 && v_3 = V_{10,10} =\frac{6 e^2}{\pi \epsilon_s} \int\limits_{0}^{\infty} 
 \frac{dq}{(\alpha q)^3}\left\{ 2 \alpha  q \cos(\alpha q)+ \left((\alpha q)^2 - 1 \right) 
\sin(\alpha q) \right\} j_1(R_1 q)j_1(R_2 q)  
\end{eqnarray}
The $\alpha -$dependence accounts for the fact that   two adjacent spheres 
($1-2$ and   $2-3$) have equal separation  $a$, whereas the far-removed pair $1$ 
and $3$ have their centers separated by distance $2a$.

\medskip
\par

For a $z-$aligned triad,
 each interaction matrix block matrix consists of only  two  different
Coulomb potential matrix elements  and, most crucially, is  diagonal. This established
fact  results in the factorization of the determinant of the  
${\cal M}_{\epsilon}$ matrix. The resulting product matrix is
 
\begin{equation}
{\mbox Det}\  \tensor{{\cal M}}_{\epsilon} = \prod_{\lambda=1}^4  m_\lambda
\label{product}
\end{equation}
with

\begin{eqnarray}
&&  m_1 = \left(\epsilon_{L=1}(R,\,\omega) - 
\frac{2 e^2}{\epsilon_s} \Pi_{L=1}(\omega) v_1(R,2a) \right)^2 \\
&&  m_2 =  \epsilon_{L=1}(R,\,\omega) - 
\frac{2 e^2}{\epsilon_s} \Pi_{L=1}(\omega) v_3(R,2a)  \\
&&  m_3 = \left( \epsilon^2_{L=1}(R,\,\omega) + 
\epsilon_{L=1}(R,\,\omega) \frac{2 e^2}{\epsilon_s}
 \Pi_{L=1}(\omega) v_1(R,2a)  - 2 \left(\frac{2 e^2}{\epsilon_s} 
\Pi_{L=1} v_1(a,R) \right)^2 \right)^2 \\
&&  m_4 = \epsilon^2_{L=1}(R,\,\omega) + \epsilon_{L=1}(R,\,\omega) 
\frac{2 e^2}{\epsilon_s} \Pi_{L=1}(\omega) v_3(R,2a)
- 2 \left(\frac{2 e^2}{\epsilon_s} \Pi_{L=1} v_3(a,R) \right)^2
\end{eqnarray}
each ``factored linear" multiplier results in one solution, either singly or doubly
degenerate, whereas each ``quadratic" part gives rise  to two different solutions.

\begin{figure}
\centering
\includegraphics[width=0.48\textwidth]{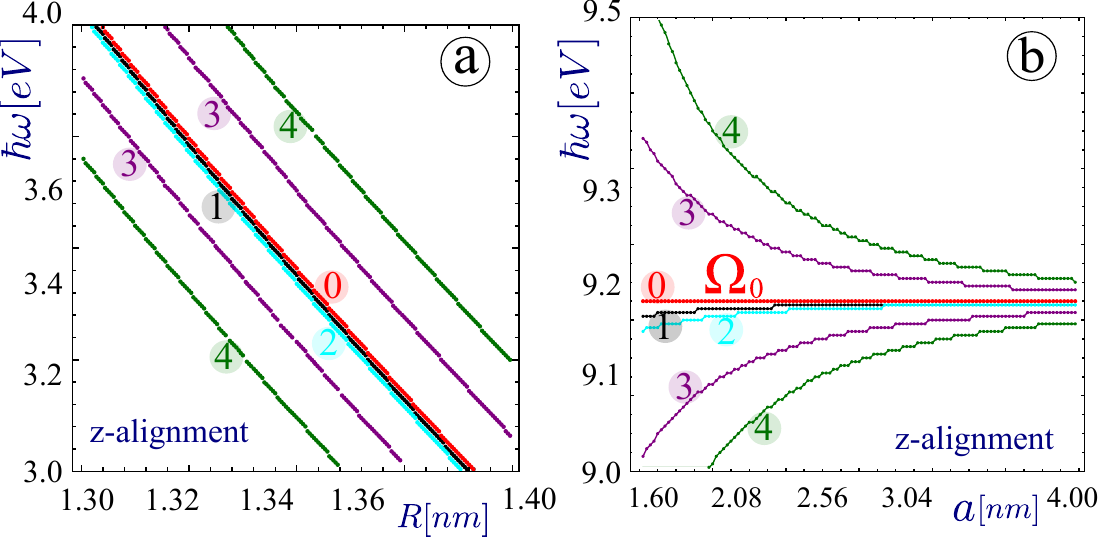}
\caption{(Color online) Exact numerical solutions for the plasmon frequency for three 
$z-$aligned identical S2DEGs with equal radii $R_1=R_2=R$ and   separation between 
them $a_1=a_2=a$. The plasmon frequency dependence on the radius of the spheres for the
range $1.3\,nm\,<\,R\,<1.4\,nm$ is shown at plot $(a)$. The separation between their 
centers is $a=2R+0.1\,nm$. Plot $(b)$ represents the frequencies of the plasmon branches
for the three spheres with $R=0.8\,nm$ as a function of the separation $a$. The solution 
for a system of non-interacting shells $\Omega_0=9.17\,eV/\hbar$ is also provided for 
reference. Since the determinantal equation may be factorized, we clearly separate for 
different solutions $1\,-\,4$, although \textit{two} of them are doubly degenerate 
($1$ and $3$), and each   of the other \textit{two} solutions $3$ and $4$ yields
 \textit{two} plasmon branches (\textit{four} in total).  
}
\label{FIG:5}
\end{figure}

\par
\medskip
For $x-$aligned shells, the potential submatrix is not diagonal, so 
similar factorization is not possible and we have

\begin{equation}
\label{elx}
 \mbb{V}_{jj'}=\Pi_{L=1}(R,\,\omega)
\left({  \begin{array}{ccc}
v_1  &  0  & v_2\\
0  &  v_3  &  0\\
v_2  &  0  & v_1
\end{array}  }\right)\ .
%\label{no-product}
\end{equation}
with the elements as follows:

\begin{eqnarray}
 && v_1 = V_{1-1,1-1} = V_{11,11} =\frac{3 e^2}{\pi \epsilon_s} \int\limits_{0}^{\infty} 
 \frac{dq}{(aq)^3}\left\{ aq \cos(aq)+\left((aq)^2 -1 \right)\sin(aq) \right\} j_1(R_1 q)j_1(R_2 q) \\
 \nonumber
 && v_2 =  V_{1-1,11} = V_{11,1-1} =\frac{3 e^2}{\pi \epsilon_s} \int\limits_{0}^{\infty} 
 \frac{dq}{(aq)^3}\left\{ 3aq \cos(aq)+\left((aq)^2 -3 \right)\sin(aq) \right\} j_1(R_1 q)j_1(R_2 q) \\
 \nonumber
 && v_3 = V_{10,10} =\frac{6 e^2}{\pi \epsilon_s} \int\limits_{0}^{\infty} 
 \frac{dq}{(aq)^3}\left\{ aq \cos(aq)+ \sin(aq) \right\} j_1(R_1 q)j_1(R_2 q) \\
\end{eqnarray}

\par
As the final step of our discussion of the potential elements,
we address the properties of the largest one, corresponding to
$L=L'=0$ and $M=M'=0$. It does not contribute to the plasmons of our
system because $\Pi_{L=0}(\omega) = 0$. Clearly, it demonstrates 
complete spherical symmetry, i.e. this is the only element, having this same  property 
for both $x-$ and $z-$ alignments. Also, due to its specific symmetry 
properties, this is the largest element. Some straightforward integration
in Eq.[\ref{eq1.51b}] leads to the following expression:

\begin{equation}
V_{00,00}(R_1,R_2,a) = \frac{3 e^2}{\pi \epsilon_s} \int\limits_{0}^{\infty} dq j_0(R_1 q) j_0 (R_2 q) j_0(a q)
\end{equation}
In the following section, we employ these results to determine 
numerically the plasmon excitations when three shells are lined up along 
or perpendicular to the axis of quantization.

%
%
%%%%%%%% \textcolor{red}{Here we stop making any corrections}

%%%%%%%%%%%%%%%%%%%%%%%%%%%%%%%%%%%%%%%%%%%%%%%%%%%%%%%%%%%%%%%%%%%%%%%%%%%%%%%%%%%

\section{Numerical Results}
\label{sec4}
As the first step for our numerical results, we investigate the Coulomb interaction 
potential matrix elements $V_{j',L',M',j,L,M}(R_j, R_j', a_{jj'})$, which are involved
in determining  the plasmon excitations of the assembly according to Eq. (\ref{eq1.51}).  
The calculation of the potential matrix elements of each interacting pair of S2DEGs is similar to our 
previous considerations \cite{2spheres,3spheres}. All non-zero potential matrix elements are given in 
Fig.\ \ref{FIG:2} for $L=1$. We note the each inter-sphere matrix element has a strong dependence
 on the distance between the shells, so that the interaction is significant for a closely packed assembly with 
$a_{jj'} \backsimeq R_j +  R_{j'}$. This behavior is similar to the single shell potential 
matrix element $\backsimeq 1/[(2L+1)R_j]$.     The magnitude of each element at a given distance 
depends on the symmetry of the quantum state, i.e. the values of $L,\,L'$, as well as $M$ and $M'$. 
The largest one corresponds to rotational symmetry when $L=L'=0$ and $M=M'=0$. However, it does not 
couple with irradiation with specific polarization and, therefore does not contribute to the 
plasmon excitation since the polarization function is identically zero for zero angular momentum.   
All   other elements are of the same order of magnitude. Existence of non-zero potential matrix
elements are given by specific selection rules,  representing an interesting contribution of our work. 
Thus, due to symmetry of the interaction, the only non-zero elements for z-alignment are those with 
$M=M'$, which is not the case for the alternative situation of $x-$alignment. Only the elements with $L=L'=1$ 
contribute to the plasmon spectrum for  both cases of shell alignment. As a result, we are left with 
 three  different non-zero interaction matrix elements in the case of $x-$alignment, while the 
shells aligned along the $z$ axis exhibit only  two of them, which is reflected in constructing 
the interaction matrices in Eq.\ (\ref{ma}). The insets in Fig. \ \ref{FIG:2}
 show the behavior of each potential matrix element for the case of unequal radii of the 
interacting shells. We conclude that there is no  pattern for such dependence, except 
that each element, either positive or negative, increases (the absolute value) as a 
function of $\rho = R_2/R_1$ for approximately equal radii $R_1 \backsimeq R_2$.

\par
\medskip
\par

We now turn to a discussion of our numerical results for the plasmons of a linear triad
 of S2DEGs  for angular momentum $L=1$.  The plasmon excitation energies of three 
 Coulomb-coupled S2DEGs are obtained by solving  
for the zeros of the determinant of the matrix in  Eq.\ (\ref{product}) when the S2DEGs are
 centered on the $z$ axis and solving the general equation (\ref{ma}) with the 
 interaction submatrix given in Eq.(\ref{elz}).
 In general, we must deal with a matrix of infinite order 
whose elements are given by the coefficients of Eq.\  (\ref{eq1.51b}) since all 
angular momenta  corresponding to $L= 1,\,2\,\cdots,\,\infty$ are coupled. However, 
for incoming light with a specific polarization, only one angular momentum quantum number 
might be needed in carrying out  the calculation to determine the plasmon frequencies 
since in general the angular momentum of light may be carried
by either orbital motion (helicity) or spin motion (circular
polarization). If a finite angular momentum of light with L = 1 is used
for incidence,  the magnetic field generated
from the induced oscillating electric dipole moment on one
sphere can couple to the induced magnetic dipole moment on
another displaced sphere.   This results in a $9 \times 9$ 
determinantal equation with  elements corresponding to $j=1,\,2$ and $3$ 
in Eq.\ (\ref{eq1.51b}) with  $L=1$ and $M=0\,\pm1$. The resulting matrix consists of thee $3 
\times 3$ diagonal blocks, related to the plasmons of each individual S2DEG and the 
remaining six off-diagonal $3 \times 3$ submatrices with $j \neq j' =1,\,2$ and $3$. 
The diagonal block could be either identical for the case of three equivalent shells,
or different, depending on the radius and the number of electrons on each   S2DEG. Each   
block is diagonal, which means that for non-interacting shells the total matrix is 
diagonal and its determinant may be easily factorized.  

\medskip
\par

Both the intra- and inter-shell Coulomb interactions contribute to the plasmon
excitation energies. First, we determine how the plasma frequencies depend on the radius of three identical
S2DEGs. Our results are presented in Fig. \ \ref{FIG:3}. First, we present
the single-particle excitation region (SPER) in the   $\omega$-$R$ plane  as being identified by finite imaginary
part of the non-interacting electron polarization function $\Pi_{L,M}(\omega)$ with the 
$R-$dependence arising  from the single-particle energies in Eq.\ (\ref{eq:sp}). The SPER, which is
also referred to as the particle-hole excitation region, shows the  $\omega$-$R$  regions where
there is natural damping of the plasmon modes.     We see by comparing our plots in Figs.\ \ref{FIG:3}(a) 
and \ref{FIG:3}(b) that  the plasmons  for a single S2DEG are  undamped, well separated from 
the SPER and their energies decrease  monotonically with increasing radius.
We present our results in Figs.\   \ref{FIG:3}(c)  through  \ref{FIG:3}(f) for interacting shells.  The two 
distinct cases of either $x-$ or $z-$alignment  clearly   lead to different solutions for 
the plasmon modes.  Whereas for $x-$alignment, there are  nine  branches, there are only  six 
 solutions for the plasmons in a $z-$aligned assembly. The corresponding $a-$dependence 
(vs. the distance between the S2DEG centers) is displayed in Figs.\ \ref{FIG:4}*a) through
\ref{FIG:4}*d).    In general, the plasmon modes do not have uniform  intensity. Specifically,
the peaks are not strong for the lower frequency modes because of their proximity   to
 the particle-hole mode region.  We found that when the separation $a$ increases, the 
plasmon branches tend to the single degenerate solution  denoted by $\Omega_0$, 
corresponding to non-interacting shells.                

\par
\medskip
\par
The mathematically simpler case when there is $z-$alignment, which allows one to factorize 
the determinantal  equation for the matrix in Eq.\ (\ref{product}),  deserves   some attention. 
 Our calculations yield each plasmon mode as a numerical solution of  the determinantal equation. 
and present our results in Fig.\ \ref{FIG:5}.  Once again, we confirm that there are  six
 distinct branches, corresponding to the four factored matrices, two of them are quadratic
 resulting in two solutions each. Two of the solutions are doubly-degenerate. The branches 
are not symmetric  in intensity with respect to the horizontal line  $\Omega_0 $. 

\par
\medskip
\par

\begin{figure}
\centering
\includegraphics[width=0.48\textwidth]{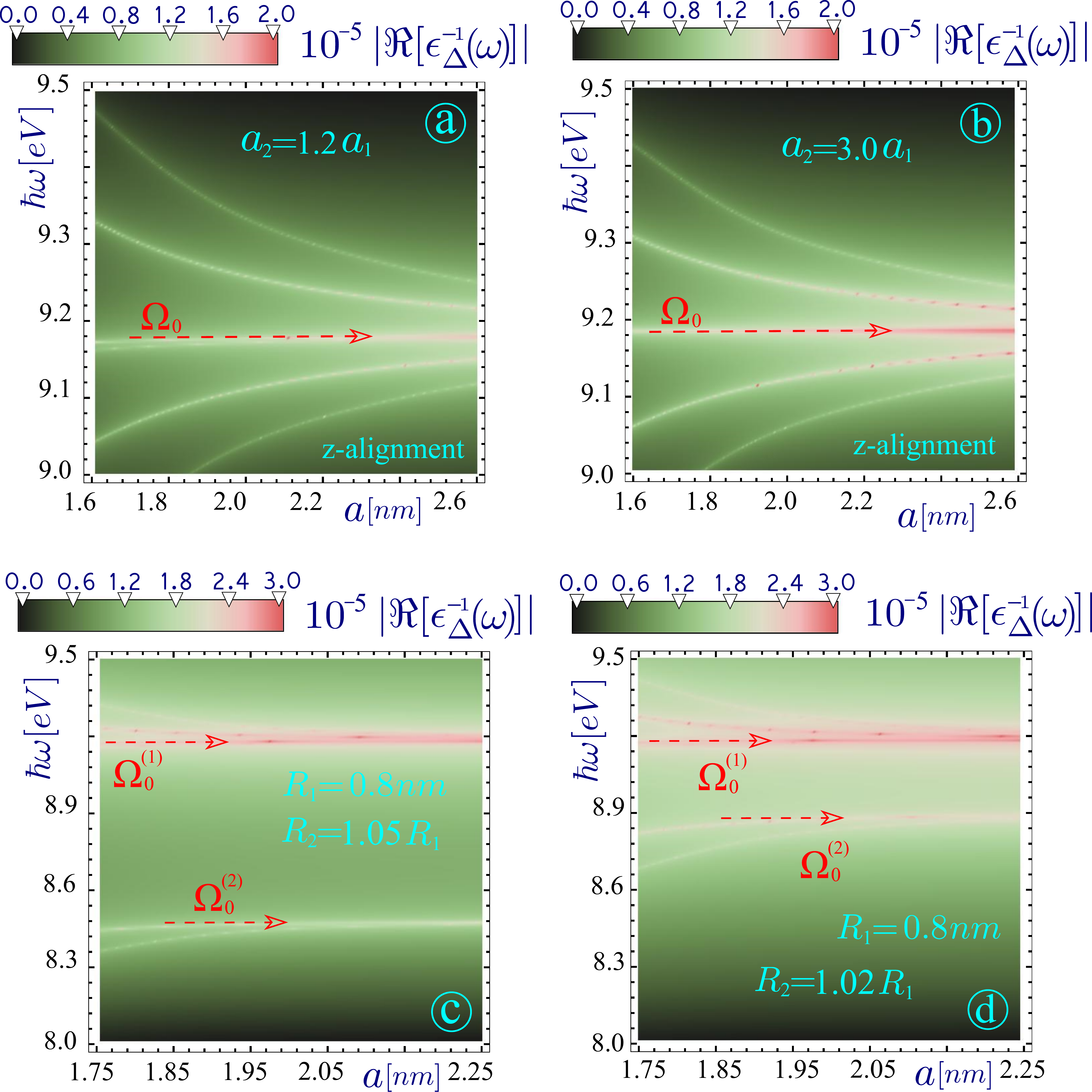}
\caption{(Color online). Plasmons for an asymmetric linear assembly of S2DEG's. 
The upper panel demonstrates the plasmon solutions for three equivalent fullerenes
with different distances between their centers: $a_2 = 1.2 \, a_1$ for plot $(a)$ 
and $a_2 = 1.2 \, a_1$ - for $(b)$. Distances are specified according to the 
schematics in Fig.\ref{FIG:1}. The range of the smaller distance $a_1 = a_{12}$
is $1.6\, - \, 2.6\, nm$ for both plots $(a)$ and $(b)$. The solution for a system
of non-interacting shells $\Omega_0=9.17\,eV/\hbar$, being the asymptotic solution,
is provided for comparison. The lower panels $(c)$ and $(d)$ demonstrate the plasmons
for three different spheres with $R2 \neq R_1 = R_3$, $R_2/R_1 = 1.05$ and $1.02$, 
correspondingly. The distances between the spheres are equal: $a_1=a_2=a$. Each plot has 
two asymptotic solutions - $\Omega_0^{(1)}=9.17\,eV/\hbar$, corresponding to $R_{1} = 
0.8 \, nm$, $\Omega_0^{(2)}=8.48\,eV/\hbar$ for $R_{2} = 0.84 \, nm$ (panel $(c)$) and 
$\Omega_0^{(1)}=8.87\,eV/\hbar$ for $R_{2} = 0.82 \, nm$ (panel $(d)$). The distance $a$
between the centers of the spheres is displayed in the range of $1.75\, - \, 2.25\, nm$
for both plots $(a)$ and $(b)$. We chose the case of z-alignment for all the plots. 
}
\label{FIG:6}
\end{figure}

So far, in this paper, we presented numerical results when the radii of the spherical 
shells are equal and also when the separations between nearest-neighbors are equal.  
Figures\ \ref{FIG:6}(a)  through \ref{FIG:6}(d) examine the behavior of the plasmon 
modes of a linear assembly with unequal nearest-neighbor separations
  as well as when the S2DEGs have different radii and chemical potentials, 
which   corresponds to more realistic situation of fullerene aggregates. First,
in this regard, we have  investigated the mathematically simpler case of  $z-$aligned shells
with equal radius $R=0.8$ nm but unequal distances between their centers. 
In this regard, there is no  symmetry of the 
interaction submatrices and we may no longer use the matrix factorization in Eq.\
(\ref{product}). Instead, we obtain a lower-
order factorization, which consists of two multipliers, each of cubic order. Therefore, 
we still have  six solutions, but there is no degeneracy of the frequencies. Once the ratio $\gamma=
a_2/a_1$ is increased, the electrostatic interaction between the nearest-neighbor
shells which we refer to as  $(2)$ and $(3)$, 
and naturally between next-nearest-neighbor shells $(1)$ and $(3)$ is decreased,  
with the plasmon frequency  of the assembly
tending to the single-shell  plasmon mode frequency $\Omega_0 =9.17, eV/\hbar$. 
The remaining plasmon branches are similar to those of the interaction between
$(1)$ and $(2)$. These branches are symmetric with respect to the asymptotic line $\Omega_0 $, as
we obtained for the case of two interacting $z$-aligned shells \cite{2spheres}.
When the distances between nearest-neighbor S2DEGs  are comparable, the plasmon modes do
not behave as $\pm v_i$, which is a new feature compared to all the previously considered
cases \cite{2spheres,3spheres}.

\medskip
\par

Finally we address the case when the radius of the middle shell $(2)$ is larger than the
radius of the other two: $R_2 > R_1 = R_3$ with $\rho = R_2/R_1 > 1$. Now we have two 
asymptotic lines $\Omega_0^{(1)}$ and $\Omega_0^{(2)}$. The larger solution 
$\Omega_0^{1}$ corresponds to the \textit{two} spheres with lower radius $R_1=R_3$, i.e. 
is double degenerate, which results in a stronger peak (see Fig.[\ref{FIG:6}](c) and (d)).
The plasmon modes are strongly asymmetric. We also conclude that only two branches have their
asymptotes along the line $\Omega_0^{(2)}$, corresponding to the middle shell $(2)$.

\section{Concluding Remarks}
\label{sec5}

In this paper, we carried out a model calculation of the plasmon  relation
for a narrow ribbon of fullerene atoms or metallic shells which were simulated
 by a linear  array of S2DEGs.  In neglecting the width of the ribbon, we
only  investigated  effects arising from the edges along the length and not the 
width of the narrow ribbon.   The  coupling between
plasmon excitations for the 2D electron gases  on the surfaces of
 an finite number of  shells is introduced using the random-phase approximation.
We demonstrated that the quantization of the energy levels for each 3DEG
on the surface of a spherical shell leads to Coulomb matrix elements depending on
the angular momentum $L$ and its component $M$. These Coulomb matrix elements represent
the inter-shell coupling of the plasmon excitations and are clearly anisotropic, depending
on the angle the line-of-centers makes with the quantization axis 
leading to  anisotropy in the plasmon coupling  with respect to the direction of the probe field.
Furthermore, we proved that the strength of the inter-shell Coulomb interaction does not only
depend on the distance between shells but also on the direction of the line joining these 
centers. We presented detailed results for the plasmon excitation frequencies 
as functions of the separation between shells and their radii. The effects of unequal separation
between nearest-neighbor shells and when the shells have unequal radii were also investigated.

\par
		
From an experimental point of view, when light with a specified  finite orbital or spin
angular momentum is incident on the ribbon, the magnetic field generated from an induced
oscillating electric dipole on any  sphere may couple to an induced magnetic dipole on 
one of the  spheres in the array where the coupling strength is determined by  their orientational  direction
either parallel or perpendicular  to the probe ${\bf E}$ field. This leads to dimerization
of pairs of  S2DEGs  confined on two separate shells.
Therefore, the spectra of the plasma excitations are different in
the cases when the quantization axis is parallel or perpendicular to
the array axis.

\medskip
\par

This work covers fundamental aspects such as anisotropy,
many-particle quantum effects and electron-plasmon interactions 
for  a novel plasmonic material.  Applications such as 
biosensors for health care,  devices for telecommunications, 
and near-field instrumentation may be explored.
We note that our  model calculations  may  not only be  applicable to  metallic particulates  but
 also have broader  implications to  fullerenes. The numerical results we derived  
demonstrate significant new information in the area of plasmonics and are
 experimentally observable..   In this connection,  there have already been
some experimental measurements showing   similar effects in nanoparticles \cite{Yang1}.
Also,   our work  has a bearing on that of
 hybridization for surface plasmons in metallic dimers \cite{2spheres,Nordlander}
in which the  plasmon frequencies  depend on whether the quantization axis
 is parallel or perpendicular to   the inter-particle axis.

%%%%%  XXXXXXXXXXXXXXXXXXXXXXXX

\acknowledgments
This research was supported by contract $\#$ FA 9453-13-1-0291 of AFRL.


\newpage
\begin{references}

\bibitem{AFS}  T. Ando, A. B. Fowler, F. Stern,
 Rev. Mod. Mod. Phys.  {\bf 54}, 437  (1982).  



\bibitem{Jain-Allen}  J. K. Jain and P. B. Allen,
\prl {\bf 54}, 2437  (1985).


\bibitem{Xiaodong}  X. Zhu, J. J. Quinn, and G. Gumbs,
Solid State Communications {\bf 75},   595 (1990).



\bibitem{Inaoka}  Takeshi Inaoka,  Surface Science \textbf{273}, 191  (1992).


\bibitem{Devreese}  J. Tempere,  I. F. Silvera,  and J. T. Devreese, \prb
\textbf{65}, 195418 (2002).


\bibitem{Yannouleas} Constantine Yannouleas, Eduard N. Bogachek,
and Uzi Landman,  \prb  \textbf{53}, 10225   (1996).



\bibitem{2spheres}  Godfrey Gumbs, Andrii Iurov,     
Antonios Balassis, and Danhong Huang,   
J. Phys.: Cond. Matt. \textbf{26}, 135601  (2014).



\bibitem{Nordlander} P. Nordlander,  C. Oubre,  E. Prodan, 
 K. Li, and M. I. Stockman,  Nano Letters, {\bf 4}, 5 (2004).


\bibitem{3spheres}   Andrii Iurov,  Godfrey Gumbs, 
Bo Gao, and Danhong Huang,
Applied Physics Letters \textbf{104},  203103 (2014).


\bibitem{infinite}    Antonios Balassis and  Godfrey Gumbs,
\prb  {\bf 90}, 075431    (2014).



\bibitem{Reviewer_1-1} Stefan A. Maier, Mark L. Brongersma, Pieter G. Kik, 
and Harry A. Atwater,
 Phys. Rev. B  \textbf{65}, 193408  (2002). 

\bibitem{aizin}   Godfrey Gumbs and G. R. A\v{\i}zin,   
\prb     {\bf 65}, 195407-1 - 195407-6 (2002).


\bibitem{mcneish} Tibab McNeish,   Godfrey Gumbs, and A. Balassis, \prb
{\bf 77}, 235440 (2008).

\bibitem{Reviewer_1-2}  M. L. Brongersma, J. W. Hartman, and 
H. A. Atwater, Phys. Rev.   B \textbf{62}, R16356 (2000).



\bibitem{n21} M. S. Dresselhaus, G. Dresselhaus, P. C. Eklund,,
  {\em Science of Fullerenes and Carbon Nanotubes: Their Properties
  and Applications\/},    (Academic Press, 1995).

\bibitem{n22} Li-Feng Wang and Quan-Shui Zheng, Appl. Phys. Lett. \textbf{90}, 153113 (2007).

\bibitem{n23} C. W. Chiu, F. L. Shyu, M. F.  Lin,   Godfrey Gumbs, and
Oleksiy  Roslyak, J. Phys. Soc. Jpn. \textbf{81}, 104703 (2012).

\bibitem{n24} Michael K. Yakes, Daniel Gunlycke, Joseph L. Tedesco, PaulM. Campbell, 
Rachael L.Myers-Ward, Charles R. Eddy, Jr., D. Kurt Gaskill, Paul E. Sheehan, and Arnaldo R.
Laracuente,  Nano Lett.  { \bf 10}, 15591562 (2010).



\bibitem{n42} R Singhal, D C Agarwal, Y K Mishra, F Singh, J C Pivin, 
R. Chandra and D K Avasthi, J. Phys. D: Appl. Phys. { \bf 42}  155103 (2009).



\bibitem{Yang1} Yang, Shu-Chun and Kobori, Hiromu and He, Chieh-Lun and Lin, 
Meng-Hsien and Chen, Hung-Ying and Li,  Cuncheng and Kanehara, Masayuki and 
Teranishi, Toshiharu and Gwo, Shangjr, Nano Letters, {\bf 10}, 2 (2010).

\end{references}
\end{document}